\documentclass[twocolumn,aps,prl,showpacs]{revtex4}

\usepackage{graphicx}
\usepackage{bm}
\DeclareGraphicsExtensions{ps,eps,epsf}

\newcommand{ \rucuo }{\mbox{RuSr$_2$GdCu$_2$O$_8$}}

\newcommand{ \grad }{\mbox{$^{\circ}$}}

\newcommand{ \ttg }{\mbox{$t_{2g}$}}
\newcommand{ \eg }{\mbox{$e_{g}$}}

\newcommand{ \ltwo }{\mbox{$ L_{2}$}}

\begin{document}

\title{Magnetic structure of \rucuo\ determined by resonant x-ray diffraction}
\author{B. Bohnenbuck$^1$, I. Zegkinoglou$^1$,  J.~Strempfer$^2$, C.S.~Nelson$^3$, H.-H.~Wu$^{4,5}$, C.~Sch\"u{\ss}ler-Langeheine$^4$, M.~Reehuis$^{1,6}$, E.~Schierle$^6$,  Ph.~Leininger$^1$, T. Herrmannsd\"orfer$^7$, J.C.~Lang$^8$, G.~Srajer$^8$, C.T.~Lin$^1$, and B.~Keimer$^1$}
\affiliation{$^1$ Max-Planck-Institut f\"ur Festk\"orperforschung, Heisenbergstr. 1, 70569 Stuttgart, Germany}
\affiliation{$^2$ Hamburger Synchrotronstrahlungslabor HASYLAB at DESY, Notkestr. 85 , 22603 Hamburg, Germany}
\affiliation{$^3$ National Synchrotron Light Source, Brookhaven National Laboratory, Upton, New York 11973-5000, USA}
\affiliation{$^4$ $II.$ Physikalisches Institut, Universit\"{a}t zu K\"{o}ln, Z\"{u}lpicher Stra{\ss}e 77, 50937 K\"{o}ln, Germany}
\affiliation{$^5$ National Synchrotron Radiation Research Center, Hsinchu 30076, Taiwan}
\affiliation{$^6$ Helmholtz-Zentrum Berlin, Albert-Einstein-Stra{\ss}e 15, 12489 Berlin, Germany}
\affiliation{$^7$ Hochfeld-Magnetlabor Dresden (HLD),Forschungszentrum Dresden-Rossendorf, 01314 Dresden, Germany}
\affiliation{$^8$ Advanced Photon Source, Argonne National Laboratory, Argonne, Illinois 60439, USA}

\date{\today}

\begin{abstract}
X-ray diffraction with photon energies near the Ru $\ltwo$-absorption edge
was used to detect resonant reflections characteristic of a G-type superstructure in
\rucuo\ single crystals. A polarization analysis confirms that these reflections are due to magnetic order of Ru moments, and the azimuthal-angle dependence of the scattering amplitude reveals that the moments lie along a low-symmetry axis with substantial components parallel and perpendicular to the RuO$_2$ layers. Complemented by susceptibility data and a symmetry analysis of the magnetic structure, these results reconcile many of the apparently contradictory findings reported in the literature.
\end{abstract}


\maketitle

\rucuo\ (Ru1212) and related compounds with alternating RuO$_2$ and CuO$_2$ layers have attracted tremendous
scientific interest in recent years, mainly due to the microscopic coexistence of long-range
magnetic order and superconductivity \cite{Nac06,Kla08,Chu05}. With a magnetic ordering temperature $T_N = 100-150$~K and a superconducting
transition temperature of $\sim 15-50$~K, Ru1212 exhibits not only the
highest magnetic transition temperature among all magnetic
superconductors, but also the broadest coexistence range of magnetic order and
superconductivity. However, as most of the research on Ru1212 has thus far been performed on powder samples, information about the nature
of the magnetic order is limited. Neutron powder diffraction (NPD, which is complicated by the large neutron absorption cross section of Gd) has revealed two superstructure reflections below $T_N$, which indicate antiferromagnetic (AF) order of Ru moments in all three crystallographic directions (``$G$-type antiferromagnetism") \cite{Lyn00,Jor01}. While a magnetic structure refinement could not be performed, the NPD data suggested a magnetic moment direction along the $c$-axis (perpendicular to the RuO$_2$ layers). Magnetization \cite{Ber99, Wil00}, ferromagnetic resonance (FMR) \cite{But01}, and nuclear magnetic resonance (NMR) data \cite{Tok01, Kum01, Han06}, on the other hand, have been interpreted in terms of a state in which ferromagnetic RuO$_2$ layers with in-plane moment orientation are antiferromagnetically stacked along $c$. The net ferromagnetic exchange field in the CuO$_2$ layers implied by the latter scenario would actuate the intimate coupling between ferromagnetism and $d$-wave superconductivity that has motivated much of the work on Ru1212 \cite{Nac06, Kla08, Chu05}. The superstructure reflections in the NPD experiment might then be understood as manifestations of Ru$^{4+}$-Ru$^{5+}$ charge order, for which there is independent evidence from magnetometry, x-ray absorption, and NMR \cite{Tok01, Kum01, Han06, Liu01}.

In an attempt to resolve this controversy, we have performed a resonant x-ray diffraction (RXD) study of Ru1212. The photon energy was tuned to the $\ltwo$-absorption edge of ruthenium, where electric dipole transitions directly probe
the partially occupied Ru $4d$ electron orbitals responsible for the magnetic properties of the material.
The large resonant enhancement of the magnetic scattering
cross-section, in combination with the high
brilliance of the x-ray beam provided at third-generation synchrotron
facilities, enabled the investigation of sub-millimeter-sized
crystals of Ru1212 that are well below the size limit for neutron diffraction. The RXD data confirm the presence of the superstructure reflections observed by NPD \cite{Lyn00,Jor01}, and a polarization analysis of the scattering cross section rules out interpretations in terms of charge ordering. The azimuthal-angle dependence of the RXD cross section (as well as magnetization data taken on the same single crystals) demonstrate, however, that the magnetic moments in the $G$-type AF state are oriented along a low-symmetry crystallographic direction with a substantial in-plane component, as inferred from NMR and FMR data \cite{Tok01, Kum01, Han06, But01}. We show that this observation is also consistent with the previously reported NPD data \cite{Lyn00, Jor01}. A symmetry analysis of the magnetic structure implies a ferromagnetic component of the RuO$_2$ layer magnetization that alternates along $c$. Evidence for this component is provided by the macroscopic susceptibility. These observations reconcile a variety of apparently contradictory reports in the literature.

The crystal structure of Ru1212 is approximately tetragonal with room-temperature lattice
parameters $a=b=3.836$~\AA\ and $c=11.563$~\AA\ \cite{Chm00}. The electronically active units are alternating RuO$_2$ layers and CuO$_2$ bilayers
that extend parallel to the crystallographic $ab$ plane. Although subtle orthorhombic distortions have been reported in the literature \cite{Mar}, we index the wave-vector components $(h\;k\;l)$ in the tetragonal
space-group $P4/mmm$, except where noted otherwise. The investigated samples were single
crystals with typical sizes $100\times~100\times~50$~$\mu m^3$, grown by the
self-flux method as explained elsewhere \cite{Lin01}. They were picked out
of a large polycrystalline piece that also contained other phases (byproducts
of the growth procedure), and crystallographically identified with a laboratory
x-ray generator.  Magnetization measurements (inset in Fig. 2) revealed a magnetic ordering temperature of 102 K, in good agreement with prior single-crystal data \cite{Nac06}, but lower than that of most polycrystalline samples reported in the literature \cite{Nac06, Kla08, Chu05}. On the other hand, the superconducting transition temperature $T_c = 45$ K, also revealed by the magnetization measurements, is higher than that of typical Ru1212 powders. These differences probably reflect variations of the distribution of Ru, Cu, or O ions with the synthesis conditions.
The RXD experiments were conducted at beam line 4ID-D of the Advanced Photon Source
at Argonne National Laboratory and at beam line KMC1 of the BESSY synchrotron in Berlin, Germany.
At 4ID-D, the sample was enclosed in a
closed-cycle cryostat capable of reaching temperatures between 10 and 350~K, which was mounted on an eight-circle diffractometer with a vertical scattering geometry. To minimize absorption effects, the flight path was either kept in vacuum or in  He-atmosphere, and the number of Be windows was minimized. The scattered signal was detected with a NaI scintillation detector and the polarization analysis was carried out with a Si $(1\; 1\; 1)$ crystal. At KMC1, we used a UHV two-circle diffractometer with a horizonal scattering geometry designed at the Freie Universit\"at Berlin.
The sample was mounted on a copper goniometer head that was attached to the cryostat, allowing a manual rotation of the sample about the scattering vector. Sample temperatures as low as 16~K were reached with this setup.

\begin{figure}[htb]
\includegraphics[width=8.5cm]{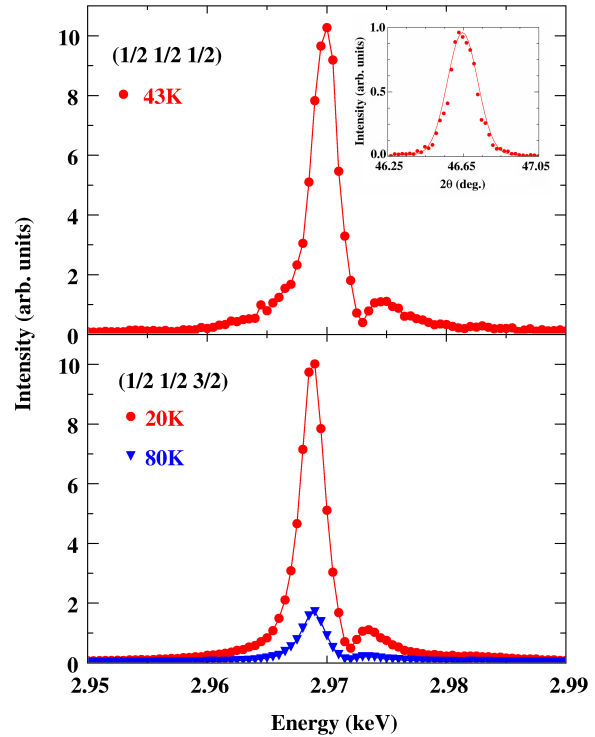}
\caption{Energy dependence of the scattered intensity at the reflections $(\frac{1}{2}\;
\frac{1}{2}\;\frac{1}{2})$ and $(\frac{1}{2}\;
\frac{1}{2}\;\frac{3}{2})$ near the Ru $\ltwo$-absorption edge. The energy profiles are not
corrected for absorption. The inset shows a typical longitudinal reciprocal space scan at reflection $(\frac{1}{2}\;
\frac{1}{2}\;\frac{1}{2})$ taken with photon energy 2.968 keV.}
  \label{fig:escans}
\end{figure}

\begin{figure}[!t]
\includegraphics[height=8cm]{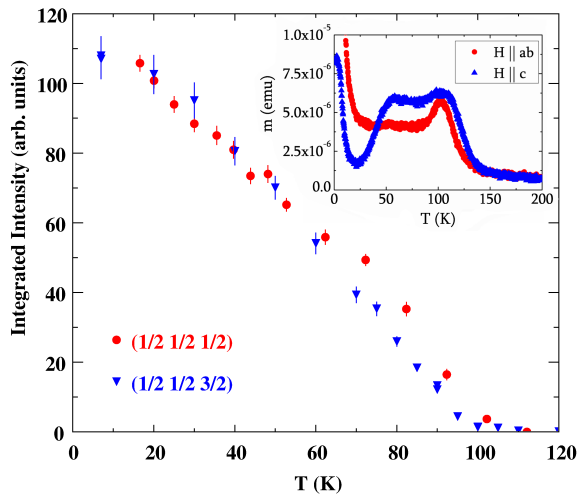}
\caption{Temperature dependence of the integrated intensity
of the magnetic reflections $(\frac{1}{2}\;\frac{1}{2}\;\frac{1}{2})$ and $(\frac{1}{2}\;\frac{1}{2}\;\frac{3}{2})$.
The N\'{e}el temperature of about 102~K agrees with the one found by field cooled magnetization measurements shown in the inset, which were carried out on single crystals at 100~Oe. }
  \label{fig:tdep}
\end{figure}

Fig. \ref{fig:escans} shows the energy dependence of the intensity of the reflections $(\frac{1}{2}\;
\frac{1}{2}\;\frac{1}{2})$ and $(\frac{1}{2}\;
\frac{1}{2}\;\frac{3}{2})$. For both reflections, a large resonant enhancement of the magnetic scattering cross-section is observed at the $\ltwo$-absorption edge. This originates from electric dipole transitions from the $2p$ core level
directly into the partly occupied $4d$~\ttg\ orbitals. A second, weaker
resonant peak approximately 4 eV above the absorption edge is
probably due to electric dipole transitions into the unoccupied $4d$~\eg\
orbitals, as previously observed in RXD experiments on Ca$_2$RuO$_4$ \cite{Zeg05}. No off-resonant scattering was observed above background. A lower bound of 500~\AA\ on the magnetic domain size in the RuO$_2$ planes was inferred from the half width at half maximum of the longitudinal reciprocal-space scan shown in the inset of Fig. 1. The intensity of the resonant reflections vanishes above the N\'{e}el temperature of 102 K (Fig. \ref{fig:tdep}), in good agreement with the magnetization data (inset in Fig. 2).

The observation of resonant magnetic reflections at reciprocal space position $(\frac{1}{2}\;\frac{1}{2}\;\frac{1}{2})$ and $(\frac{1}{2}\;\frac{1}{2}\;\frac{3}{2})$ indicates a doubling of the unit cell along all three crystallographic directions, which reflects a modulation of either the magnetization density or the charge density of the Ru valence electrons. In order to discriminate between these two scenarios, we have analyzed the polarization of the scattered signal at the $(\frac{1}{2}\;\frac{1}{2}\;\frac{1}{2})$ reflection, which was measured with an incident photon polarization perpendicular to the scattering plane.
The results show that the intensity of the superstructure reflections originates entirely from scattering events in which the photon polarization is rotated ($\sigma \rightarrow \pi'$, where $\sigma/\sigma'$ and $\pi/\pi'$ denote the polarization of the incident/scattered x-ray beam perpendicular and parallel to the diffraction plane, respectively). This confirms the interpretation of the NPD data in terms of $G$-type magnetic order \cite{Lyn00,Jor01}. As no intensity above background was detected in the $\sigma \rightarrow \sigma'$ channel, we can rule out models \cite{But01,Han06} according to which the superstructure reflections originate from charge order \cite{footnote2}.

\begin{figure}[t]
\includegraphics[height=8cm]{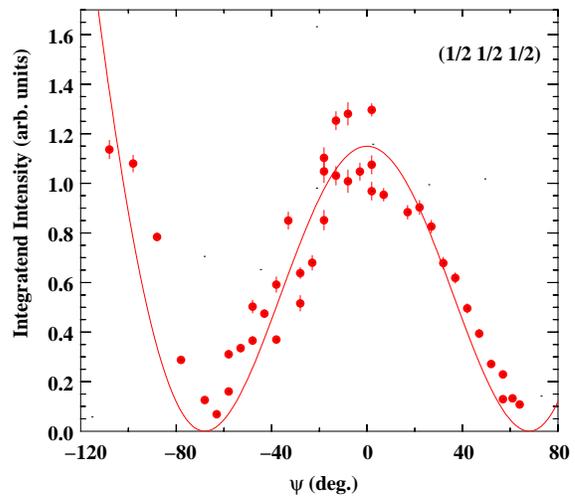}
\caption{Azimuthal dependence of the integrated scattering intensity at reflection
$(\frac{1}{2}\; \frac{1}{2}\; \frac{1}{2})$ at $T=43$~K, where $\psi=0$ is defined in such a way that $(1\; 0\; 2)$ lies in the diffraction plane. The solid line is the result of a fit to a
theoretical expression for the resonant electric dipole scattering intensity, as explained in the text.}
  \label{fig:psidep}
\end{figure}

In order to determine the direction of the magnetic moments, we rotated the sample around the scattering vector and measured the azimuthal-angle dependence of the scattered intensity at the $(\frac{1}{2}\;\frac{1}{2}\;\frac{1}{2})$ reflection (Fig. \ref{fig:psidep}). Assuming a collinear AF structure \cite{footnote1}, the results were fitted to an expression derived by Hill and McMorrow \cite{Hil96}:
\begin{equation}
\label{eq:scat_length}
I^{\sigma \rightarrow \pi'}_{(\frac{1}{2}\;\frac{1}{2}\;\frac{1}{2})} \propto
\left|  \sin\alpha \, \cos\theta \, \cos(\psi -\psi_0) + \cos\alpha \, \sin\theta \right|^2
\end{equation}

Here, $\theta$ is the Bragg angle, $\psi$ the azimuthal angle,
and $\alpha$ the angle between the magnetic moment and the scattering vector. The best fit was obtained with $\alpha = 49.0 \pm 1.1^\circ$. An additional constraint on the moment direction is provided by the phase, $\psi_0$, of the intensity modulation as a function of $\psi$, whose maximum is determined by the condition that the magnetic moment lies in the scattering plane. In our experiment, the maximum intensity was observed when the crystallographic $c$-axis was $53\grad$ off the scattering plane. Thus, the magnetic moment direction subtends an angle of $\alpha =49\grad$ with $(1\;1\;1)$ and of $53.8\grad$ with $(0\;0\;1)$. This corresponds approximately
to the $(1\;0\;2)$ direction in reciprocal space: $\angle\left((102),(111)\right)=45.6^\circ$ and $\angle\left((102),(001)\right)=56^\circ$.
Therefore, $(1\;0\;2)$ can be considered as the approximate direction of the magnetic
moment. This conclusion is supported by the azimuthal dependence measured at the second magnetic reflection $(\frac{1}{2}\;\frac{1}{2}\;\frac{3}{2})$ (not shown here), which also exhibits maximum intensity when $(1\;0\;2)$ lies in the scattering plane.

The direction of the magnetic moment inferred from our RXD data is consistent with the macroscopic susceptibility, which is reduced below the N\'{e}el temperature when the magnetic field is applied both along and perpendicular to the $c$-axis  (inset in Fig. 2).
It is also interesting to compare our results to those of the NPD experiments \cite{Lyn00}. Based on the intensity ratio of the $(\frac{1}{2}\;
\frac{1}{2}\;\frac{1}{2})$ and $(\frac{1}{2}\;\frac{1}{2}\;\frac{3}{2})$ reflections, these experiments had led to the tentative conclusion that the magnetic moments are oriented along the $c$-axis. By coincidence, this ratio happens to be identical for
the moment direction inferred from our data, so that both experiments are fully consistent. At the same time, the large in-plane component of the sublattice magnetization confirms conclusions from FMR and NMR experiments \cite{Tok01, Kum01, Han06, But01}.

In order to further assess the implications of our data, we have performed a representation analysis \cite{Ber68} of the magnetic structures compatible with the space group $Pbam$ resulting from a recent crystallographic study \cite{Mar}. In this setting, the unit cell is doubled and 45$^\circ$ rotated in the $ab$-plane with respect to the tetragonal ($P4/mmm$) unit cell, as a consequence of a staggered rotation of the RuO$_6$ octahedra around the $c$-axis.
The basis functions resulting from this analysis are $[-, -, Fz]$, $[-, -, Az]$, $[Ax, Fy, -]$ and $[Fx, Ay, -]$, where $F$ and $A$ denote parallel and antiparallel alignment of the Ru moments in the Wyckoff position 2a ($(0\;0\;0)$ and $(\frac{1}{2}\;\frac{1}{2}\;0)$) of $Pbam$, respectively.
In contrast to most magnetic insulators whose spin arrangements are described by a single irreducible representation, a description of the observed magnetic structure of Ru1212 requires a combination of irreducible representations: $[-, -, Az]$ in combination with $[Ax, Fy, -]$ and/or $[Fx, Ay, -]$. (The latter two possibilities cannot be distinguished because our crystals are composed of two orthorhombic twin domains. We cannot rule out a small admixture of $[-, -, Fz]$.) This may reflect structural distortions beyond those reported in the literature, or terms in the spin Hamiltonian of order higher than the usual bilinear exchange coupling. In Ru1212, such terms may arise from charge and/or orbital fluctuations in the RuO$_2$ layers, or from proximity to the highly conducting CuO$_2$ bilayers. Note that a similar effect was recently observed in insulating vanadates and tentatively attributed to orbital fluctuations \cite{Ree06}.

Leaving these details aside, the representation analysis reveals that a ferromagnetic in-plane component is required by symmetry to accompany the experimentally observed staggered component of the magnetization, as a consequence of a Dzyaloshinskii-Moriya interaction activated by the octahedral rotation pattern. The propagation vector ($(0\;0\;\frac{1}{2})$ in $Pbam$) implies an alternation of this component along the $c$-axis. This is precisely the magnetic mode inferred from FMR and NMR experiments \cite{Tok01, Kum01, Han06, But01}. While this component appears to be too small to be directly apparent in the RXD or NPD data, the pronounced upturn in the uniform susceptibility above $T_N$ may be an indirect manifestation of a ferromagnetic moment of each RuO$_2$ layer that is compensated by weak interlayer exchange interactions, as observed in other ``weak" ferromagnets such as La$_2$CuO$_4$ \cite{Thi88}. Defects such as stacking faults \cite{Leb06} (or else structural distortions beyond $Pbam$) may then induce an uncompensated ferromagnetic moment observed in some (but not all) experiments \cite{Nac06,Kla08, Chu05}.

In summary, our RXD data (in conjunction with prior crystallographic work \cite{Mar}) reconcile a variety of apparently contradictory findings on the magnetic structure of Ru1212 from different experimental probes, and thus resolve a major puzzle in the experimental literature. Further work is required to assess the influence of the surprisingly complex magnetic structure on the superconducting properties of the CuO$_2$ layers. In particular, it is conceivable that the exchange field imposed by domain boundaries of this structure contributes to the granular superconducting response reported previously \cite{Nac06, Kla08, Chu05}.

Work at Brookhaven was supported by the U.S. Department of
Energy, Division of Materials Science, under Contract No.
DE-AC02-98CH10886. Work at Cologne was supported by the DFG through SFB 608 and by the BMBF, Project 05 ES3XBA/5.  Use of the Advanced Photon Source was supported by the U. S. Department
of Energy, Office of Science, Office of Basic Energy Sciences, under
Contract No. DE-AC02-06CH11357.


\begin{thebibliography}{}
\bibitem{footnote2} In principle, a staggered ordering of Ru $t_{2g}$ orbitals might also contribute to the photon polarization rotation, as recently observed in Ca$_2$RuO$_4$ (Ref. \onlinecite{Zeg05}). However, the temperature dependence of the intensity of the superlattice reflections (Fig. 2) would imply that this order sets in at $T_N$. This is inconsistent with crystallographic data that reveal no major changes at $T_N$.
\bibitem{footnote1}  The deviation of the sublattice magnetization from $c$ cannot be explained as a consequence of canting. Neither an AF alignment along the $c$-axis with a ferromagnetic component in the $ab$-plane nor an AF spin arrangement in the $ab$-plane with canting along $c$ can reproduce both the periodicity of the azimuthal dependence and the experimentally observed angle between the $c$-axis and the maximum intensity position.
\end{thebibliography}
\end{document}